\documentclass[twocolumn,trackchanges]{aastex62}

\usepackage{url}
\usepackage{comment}
\usepackage {amsmath}
\usepackage{ulem}
\usepackage{color}
\def\dotdeg{\hbox{$.\!\!^{\circ}$}}
\def\dotarc{\hbox{$.\!\!''$}}
\received{2020 February 13}
\revised{}
\accepted{2020 December 18}
\submitjournal{ApJ}

\shorttitle{Broad-velocity-width Molecular Features in the Galactic Plane}
\shortauthors{Yokozuka et al.}
\begin{document}

\title{Broad-velocity-width Molecular Features in the Galactic Plane}

\correspondingauthor{Hiroki Yokozuka}
\email{gackt4869akai@keio.jp}

\author[0000-0002-0786-7307]{Hiroki Yokozuka}
\affil{School of Fundamental Science and Technology, Graduate School of Science and Technology, Keio University, 3-14-1 Hiyoshi, Kohoku-ku, Yokohama, Kanagawa 223-8522, Japan}

\author{Tomoharu Oka}
\affiliation{School of Fundamental Science and Technology, Graduate School of Science and Technology, Keio University, 3-14-1 Hiyoshi, Kohoku-ku, Yokohama, Kanagawa 223-8522, Japan}
\affiliation{Department of Physics, Institute of Science and Technology, Keio University, 3-14-1 Hiyoshi, Kohoku-ku, Yokohama, Kanagawa 223-8522, Japan}
\author{Shunya Takekawa}
\affiliation{Nobeyama Radio Observatory, National Astronomical Observatory of Japan 462-2 Nobeyama, Minamimaki, Minamisaku-gun, Nagano 384-1305, Japan}
\affiliation{Faculty of Engineering, Kanagawa University 3-27-1 Rokkakubashi, Kanagawa-ku, Yokohama, Kanagawa 221-8686, Japan}
\author{Yuhei Iwata}
\affiliation{School of Fundamental Science and Technology, Graduate School of Science and Technology, Keio University, 3-14-1 Hiyoshi, Kohoku-ku, Yokohama, Kanagawa 223-8522, Japan}
\author{Shiho Tsujimoto}
\affiliation{School of Fundamental Science and Technology, Graduate School of Science and Technology, Keio University, 3-14-1 Hiyoshi, Kohoku-ku, Yokohama, Kanagawa 223-8522, Japan}



\begin{abstract}
We performed a systematic search for broad-velocity-width molecular features (BVFs) in the disk part of our Galaxy by using the ${\rm CO}$ {\it J}=1--0 survey data obtained with the Nobeyama Radio Observatory 45 m telescope.\ From this search, 58 BVFs were identified.\ In comparisons with the infrared and radio continuum images, 36 BVFs appeared to have both infrared and radio continuum counterparts, and 15 of them are described as molecular outflows from young stellar objects in the literature.\ In addition, 21 BVFs have infrared counterparts only, and eight of them are described as molecular outflows in the literature.\ One BVF (CO 16.134--0.553) does not have any luminous counterpart in the other wavelengths, which suggests that it may be an analog of high-velocity compact clouds in the Galactic center.
\end{abstract}

\section{Introduction} 
\label{sec:intro}
Interstellar gas, which is an important ingredient of galaxies, fills the space in between the star system in a galaxy.\ It is composed of multiple phases and has a wide range of physical conditions.\ Stars form within the densest regions of interstellar gas, and newly born stars impart radiational and kinematical feedback to the ambient gas.\ As well as the stellar feedback, various interstellar processes such as supernova explosions and gravitational interactions with massive objects can imprint on the distribution, kinematics, and physical conditions of interstellar gas \citep{Oka01, Oka16}.\

The central 300 pc of our Galaxy is characterized by a strong concentration of warm and dense molecular gas, i.e., the central molecular zone (CMZ; \citealt{Morris 96}).\ Over time, it has become clear that the CMZ contains a number of compact ($d\!<\!\!10$ pc) clouds with extraordinary broad-velocity widths ($\Delta V\!\geq \!50 $ km s$^{-1}$), which have been named as high-velocity compact clouds (HVCCs;  e.g.,\,\citealt{Oka98, Oka07, Oka12}).\ Some of the HVCCs are considered to have been formed by either multiple supernova explosions in massive stellar clusters (\citealt{Tanaka07}; \citealt{Oka08}; \citealt{Tsuji18}) or encounters with massive compact objects (\citealt{Oka17}; \citealt{Take17, Take19a, Take19b, Take20}), although most of them do not have any other wavelength counterparts.\
\begin{figure*}[tbh]
\includegraphics[width=180 mm]{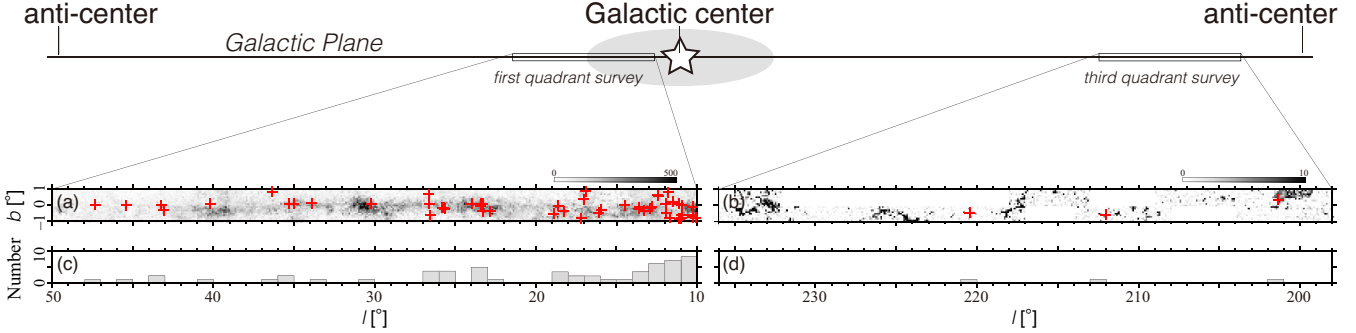}
\vspace{0.6 mm}
%\hoffset=-10cm
%\centering
\caption{(Top): schematic view of the FUGIN survey coverages.\ (Bottom): the spatial distribution of BVFs.\ Panels (a) and (b) show the $\it{l}$--$\it{b}$ distribution supposed on the maps of velocity-integrated CO emissions.\ Panels (c) and (d) show the longitudinal distributions.}
\label{fig1}
\end{figure*}
\begin{figure}[tbh]
%\vspace{0.6 mm}
\includegraphics[width=80 mm]{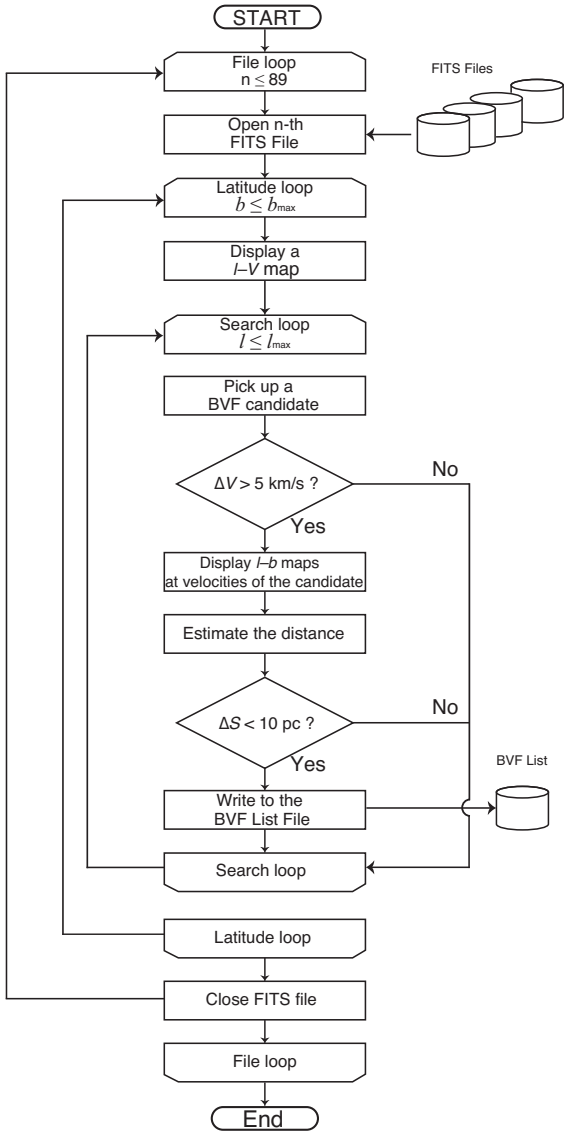}
%\centering
\caption{Flow chart of the BVF identification scheme we employed in this study.}
\label{fig2}
\end{figure}

Our group has been investigating these HVCCs for 20 yr extensively, particularly being interested in those without any apparent driving source.\ There are few reports in the Galactic disk, while $\sim \!80$ HVCCs have been identified in the CMZ so far.\ We found an extraordinary broad-velocity-width ($\Delta V\!\! \simeq\!\!100 $ km s$^{-1}$) feature in the W44 giant molecular cloud (Bullet; \citealt{Sa13}).\ This is only one example of an HVCC-like feature in the Galactic disk.\ The huge kinematic energy of the Bullet, which cannot be accounted for by the W44 supernova (SN), and its ``Y"- shape feature in the position-velocity plane can be explained by the high-velocity plunge of an isolated black hole into the dense molecular gas layer behind an SN blast wave (\citealt{Yama17}; \citealt{Nomura18}).\

In order to examine the universality of HVCCs in the Milky way, we embarked on systematic searches for broad-velocity-width molecular features (BVFs) with compact appearances in the disk part of our Galaxy.\ This project also aims to perform an unbiased search for compact BVFs, which may have been accelerated by celestial objects of well-known populations, by using a recent high-resolution CO survey of the Galactic plane.\ In this paper, we present a catalog of BVFs identified in the FUGIN ${\rm CO}$ {\it J}=1--0 Galactic plane survey data\ \citep{2017PASJ...69...78U}.

\section{Data and Identification Scheme} \label{sec:style}
\subsection{Data}
We used the CO Galactic plane survey data obtained in the FOur-beam REceiver System on the 45 m telescope (FOREST) unbiased Galactic imaging survey with the Nobeyama 45 m telescope (FUGIN) project (\citealt{2017PASJ...69...78U}).\ The FUGIN project used the multi-beam FOREST receiver installed on the Nobeyama Radio Observatory (NRO) 45 m telescope from 2014 to 2017.\ This survey achieved the highest angular resolution to date ($\sim\!\hspace{-0.1 mm}20''$) for the Galactic plane survey in the {\it J}=1--0 lines of CO and isotopologues.\ The half-power beamwidth of the telescope was $\sim\!\!\hspace{0.01 mm}14''$ at 115 GHz.\ The SAM45 spectrometer was used in the 244.14 kHz frequency resolution mode.\ The system noise temperatures ($T_{\rm sys}$) were $\sim\!\!\hspace{0.25 mm}250\rm\ K$ for the ${\rm^{12}CO}$, and $\sim\!\!\hspace{0.25 mm}150\rm\ K$ for the ${\rm^{13}CO}$ observations, respectively.\ To obtain the main beam temperature ($T_{\rm MB}$), the antenna temperatures (${T_{\rm A}^*}$) were converted by using the main beam efficiency ($\eta_{\rm MB}$) of 0.43 for ${\rm^{12}CO}$ and 0.45 for ${\rm^{13}CO}$.

\begin{deluxetable*}{ccccccccccc}
\tablecaption{Galactic disk broad-velocity-width molecular features catalog \label{chartable}}
\tablewidth{700pt}
\tabletypesize{\scriptsize}
\tablehead{
\colhead{Name} & \colhead{$l$} & 
\colhead{$b$} & \colhead{$\textit V_{\rm{par}}$} & 
\colhead{$\Delta l$} & \colhead{$\Delta b$} & 
\colhead{$\Delta V^\pm$} & \colhead{$\textit T_{\rm{peak}}$} & 
\colhead{${\rm^{13}CO}$/${\rm^{12}CO}$} & \colhead{Type} & \colhead{Reference} \\ 
\colhead{(1)} & \colhead{(2)} & \colhead{(3)} & \colhead{(4)} & 
\colhead{(5)} & \colhead{(6)} & \colhead{(7)} &
\colhead{(8)} & \colhead{(9)} & \colhead{(10)}& \colhead{(11)}
} 
\startdata
CO 10.054$+$0.063 & 10.054 & 0.063 & 42 & 0.038 & 0.042 & 9 ($+$) & 10.4 & 0.18 & IR &\nodata  \\
CO 10.245$-$0.385 &10.245 & --0.385 & 12 & 0.040 &0.043 & 7 ($-$)&21.1 & 0.13 &IR &\nodata \\
CO 10.256$-$0.755 & 10.256  &--0.755 & 30 & 0.036 & 0.047 &7 ($+$) &11.6&0.16& I & \nodata \\
CO 10.259$-$0.782 & 10.259  & --0.782 & 30 &0.036 & 0.047 &6 ($+$) &15.2 & 0.11 &IO& a \\
CO 10.293$-$0.150 &10.293 & --0.150 &12 & 0.023 &0.029 &22 ($\pm$) &33.7 & 0.23 &IRO&b  \\
CO 10.621$-$0.381 & 10.621  & --0.381 & --2 & 0.041&0.040 &14 ($\pm$)  & 46.2 &0.19 & IRO& c \\
CO 10.957$-$0.760 &10.957&--0.760& 27 & 0.021 &0.032 & 10 ($-$)  &10.8& 0.17 & I & \nodata \\
CO 10.999$-$0.902 & 10.999 &--0.902 &25 &0.0097 & 0.0093 &11 ($\pm$) &12.2 & 0.27  & I& \nodata \\
CO 11.105$-$0.983 & 11.105 & --0.983 & 26 & 0.010  &0.011 & 5 ($-$) &7.25 & 0.28 &I& \nodata \\
CO 11.113$-$0.894& 11.113 &--0.894 &26 &0.036 &0.019 & 5 ($+$)  & 6.31 & 0.17 &I & \nodata \\
CO 11.509$-$0.027 & 11.509 & --0.027 &33& 0.010& 0.016 & 5 ($-$)  &5.24 &0.29 & IR & \nodata \\
CO 11.771$-$0.075& 11.771 & --0.075 & 62 & 0.017 & 0.021 &6 ($\pm$)& 21.4 &0.18&IRO&c \\
CO 11.808+0.833 &11.808 & 0.833 & 25 & 0.021 & 0.020 & 10 ($\pm$) & 30.4 & 0.23 & IO &d \\
CO 11.919$-$0.603 & 11.919 & --0.603 & 35 & 0.019 &0.023 &10 ($+$) & 11.7 &0.12 & IRO&e \\
CO 11.949$-$0.039 & 11.949 & --0.039 & 44& 0.033 &0.023 &8 ($-$)  &10.6 & 0.28 &IRO & c \\
CO 12.414+0.496 & 12.414 & 0.496 &18 & 0.029 & 0.019 & 6 ($-$) & 20.8 & 0.29 & IO&f \\
CO 12.433+0.505 &12.433 & 0.505 & 20 &0.033 &0.037 & 5 ($+$) & 16.7 &0.13  & IO &f\\
CO 12.796$-$0.204 &12.796 & --0.204 & 37& 0.036 & 0.050 & 10 ($+$) & 26.7 &0.23 &IRO &g\\
CO 12.859$-$0.203 &12.859 & --0.203& 58 & 0.040 & 0.040 & 8 ($+$) & 12.5 & 0.11 & IR & \nodata\\ 
CO 12.907$-$0.260  & 12.907 & --0.260 & 37 & 0.041 & 0.026 & 13 ($+$)  & 18.6 & 0.28 &IRO & b \\
CO 12.927$-$0.204& 12.927 & --0.204 & 35 & 0.017 & 0.024 & 6 ($-$)  & 15.1 &0.23& IR & \nodata \\
CO 13.367$-$0.378 &13.367 &--0.378 & 17 &0.021& 0.017 & 7 ($-$) &12.0&0.12 & I & \nodata \\
CO 13.461$+$0.241 &13.461 & 0.241&17 &0.032 & 0.052 & 6 ($-$)  & 11.1& 0.13 &IR & \nodata \\
CO 13.968$-$0.388 & 13.968 & --0.388 &20 &0.059 & 0.080 & 10 ($-$)  & 23.1 &0.10 & IRO & c \\
CO 14.331$-$0.646 & 14.331 &--0.646 & 22 & 0.018 & 0.016&18 ($\pm$)  & 17.1 & 0.26 &IRO & b \\
CO 15.946$-$0.386 &15.946& --0.386 & 42 & 0.015 & 0.022&7 ($+$)  & 12.6& 0.28 &I & \nodata \\
CO 16.134$-$0.553 &16.134 & --0.553 &45 & 0.034& 0.056 &25 ($+$) & 11.5 & 0.11 &  \\
CO 16.932+0.975& 16.932 & 0.975 & 20 & 0.040 & 0.037 &8 ($-$) & 15.9 &0.17 &IR & \nodata \\
CO 17.033+0.302& 17.033 & 0.302 & 23 &0.038 & 0.020 & 6 ($+$) &6.78 & 0.23 & I & \nodata \\
CO 17.625$+$0.165& 17.625 & 0.165 &21 & 0.020 & 0.029 & 6 ($-$)  &17.7&0.10 &IO & h \\
CO 18.298$-$0.398 & 18.298& --0.398 &33& 0.052 & 0.059 &15 ($\pm$)  & 16.7 & 0.13 &IR & \nodata \\
CO 18.652$-$0.062& 18.652 & --0.062 &45& 0.037 & 0.034 &6 ($+$)  & 15.3 & 0.15 &IRO & c \\
CO 18.903$-$0.676& 18.903 & --0.676 & 65 & 0.030 & 0.027 &10 ($+$)  & 7.43& 0.12 & IR & \nodata \\
CO 22.838$-$0.476 & 22.838 & --0.476 & 80 & 0.080 & 0.072 &20 ($+$)  & 14.1 & 0.12 &IR& \nodata\\
CO 23.342$-$0.429  & 23.342& --0.429& 65 &0.077 & 0.090 & 10 ($-$) & 12.4 &0.24 &IR&\nodata \\
CO 23.382$-$0.111& 23.382 & --0.111&98 &0.019 & 0.015 &7 ($+$)& 6.59 &0.26& IR& \nodata \\
CO 23.438$-$0.181 & 23.438 & --0.181 &105 & 0.013 & 0.011 &15 ($+$)  & 23.1 & 0.20 & IRO& i \\
CO 23.522$-$0.035 & 23.522 & --0.035 & 85 &0.040 & 0.012 & 12 ($\pm$) &11.5& 0.26 & IR &\nodata \\
CO 23.706+0.168 & 23.706& 0.168 & 115 &0.047&0.043 &6 ($+$) & 17.5 &0.18 &IO & j \\
CO 25.642$-$0.350 &25.642 & -0.350 & 91 & 0.022 & 0.038 & 9 ($-$) &10.4 &0.14 &IR& \nodata \\
CO 25.731$-$0.199 & 25.731 & --0.199 & 93 & 0.032 & 0.036 &9 ($-$)  &10.0 &0.10 &IR &\nodata \\
CO 25.834$-$0.233 & 25.834& --0.233 & 93 &0.042 & 0.037 &10 ($-$) & 10.5 & 0.10 &IR & \nodata \\
CO 26.561$-$0.729 & 26.561 & --0.729 &62 & 0.035 & 0.041& 28 ($\pm$) &14.6 & 0.10& I & \nodata \\
CO 26.646$-$0.825 &26.646 & --0.825 & 62 & 0.025 & 0.024 & 8 ($-$)  & 5.75& 0.13 & I & \nodata \\
CO 26.658+0.605 &26.658 &0.605& 86 & 0.024 & 0.021 & 7 ($-$) & 11.4 & 0.19 &IR & \nodata \\
CO 30.208$-$0.095 & 30.208 & --0.095 & 115 & 0.032 & 0.038 & 6 ($+$) &13.1 &0.10 & IR & \nodata \\
CO 33.909+0.090& 33.909 & 0.090 & 108 & 0.032 & 0.036 & 17 ($\pm$)  & 20.2 & 0.35 &IRO & c \\
CO 35.018+0.010 &35.018& 0.010 &75 & 0.034 & 0.026& 9 ($-$)  & 11.1 & 0.12 &IR & \nodata \\
CO 35.291+0.017 &35.291 & 0.017 &92 & 0.013 & 0.015 & 10 ($+$) & 7.91 & 0.10 & IR& \nodata \\
CO 36.386+0.786 & 36.386 &0.786 &72 & 0.023& 0.032 & 9 ($-$)  & 12.3 & 0.34 &I& \nodata \\
CO 40.274$-$0.230 &40.274 & --0.230 &74 & 0.036 & 0.022 & 33 ($\pm$) &15.1 &0.23& IO & k\\
CO 43.159$-$0.029 & 43.159& --0.029 &11 & 0.046 &0.057 & 12 ($+$) &19.2 & 0.10 & IRO & l \\
CO 43.169+0.010 & 43.169 & 0.010 & 13&0.050 &0.053  &10 ($-$) & 45.9& 0.19 & IRO & c \\
CO 45.469+0.040 & 45.469 & 0.040 & 60 & 0.010 & 0.011 & 22 ($\pm$)  & 16.3 & 0.32 &IRO &m  \\
CO 47.394+0.096 &47.394 & 0.096 & 46 & 0.014 &0.023 & 7 ($+$) & 5.28 &0.10&I& \nodata \\
CO 201.339+0.288 & 201.339 & 0.288 & --1& 0.024 &0.030 &12 ($\pm$) & 11.3 & 0.16 & IR& \nodata \\
CO 212.064$-$0.739 & 212.064 & --0.739 &45 &0.019 &0.020 &10 ($\pm$)  &14.2 & 0.18&IO&j \\
CO 220.454$-$0.613 & 220.454 & --0.613 &28 &0.041  &0.045 & 12 ($\pm$) & 13.2 & 0.22&I&\nodata \\
\enddata
\tablecomments{Column (1): name, Columns (2)--(3): peak position ($l, b$) in degrees, Column (4): velocity of the parent molecular cloud in km s$^{-1}$, Columns (5)--(6): sizes along the $l$ and $b$ axes in degrees, respectively, Column (7): velocity width in km s$^{-1}$, +: positive velocity wing, $-$: negative velocity wing, Column (8): peak temperature in K, Column (9): ${\rm^{13}CO}$/${\rm^{12}CO}$ intensity ratio, Column (10): I: counterpart in infrared maps, R: counterpart in 10 GHz continuum maps, O: literature as molecular outflow, Column (11): references ; a: \citealt{Fro15}, b: \citealt{Cyagano2008}, c: \citealt{Li2018}, d: \citealt{Chen16}, e: \citealt{2017ApJ...729..124C}, f: \citealt{2019MNRAS.485.1775I}, g: \citealt{2012AZh...56..731C}, h: \citealt{Maud18}, i: \citealt{2011MNRAS.415L..49C}, j: \citealt{Maud15}, k: \citealt{Lee13}, l: \citealt{Yang2018}, m: \citealt{2019ApJ...886L...4Z}.}
\label{table1}
\end{deluxetable*}

\begin{deluxetable*}{ccccccccccc}
\tablecaption{\textbf{Physical parameters of BVFs} \label{chartable}}
\tablewidth{700pt}
\tabletypesize{\scriptsize}
\tablehead{
\colhead{Name} & \colhead{$D$} & \colhead{$S$} & \colhead{$\sigma_{\rm{V}}$}&\colhead{$L_{\rm{CO}}$}&
\colhead{$M$} &\colhead{$M_{\rm{VT}}$}& \colhead{log\,$E_{\rm{kin}}$} & 
\colhead{$t_{\rm{dyn}}$} & \colhead{$P_{\rm{kin}}$} & \colhead{$L_{\rm{IR}}$}\\ 
\colhead{} & \colhead{(kpc)} & \colhead{(pc)}&\colhead{(km s$^{-1}$)}& \colhead{($10^{2}\ \rm{K}\ \rm{km\ s^{-1}}\ pc^{2}$)} &\colhead{($10^{2}M_\odot$)} &\colhead{($10^{4}M_\odot$)}& \colhead{}&
\colhead{($10^{5}$yr)} & \colhead{($L_\odot$)} & \colhead{($10^{3}L_\odot$)} 
} 
\startdata
CO 10.054$+$0.063 & 4.7 &1.5 &3.2&0.32& 1.7 &3.1& $46.71$ & $4.6$ &0.94& $0.98$ \\
CO 10.245$-$0.385 &3.8 &1.3 &2.5&0.61& 3.2 &1.6& $46.78$ & $5.1$ & 0.97 & $2.5$ \\
CO 10.256$-$0.755 & 3.6 &1.2  &2.5&0.23& 1.2 &1.5& $46.35$ & $4.7$ & 0.40 & $0.31$ \\
CO 10.259$-$0.782 & 3.7 &1.2 &2.2&0.21& 1.1 &1.2& $46.20$ & $5.3$ & 0.25 & $0.82$ \\
CO 10.293$-$0.150 &1.8 &0.38 &7.9&0.67& 3.5  &4.7& $46.81$ & $1.5$ &3.6& $9.4$ \\
CO 10.621$-$0.381 & 4.9 &1.7&5.0&8.7& 45&8.5& $47.81$ & $3.6$&15 &$1.7\!\times\!10^{2}$ \\
CO 10.957$-$0.760 &3.4 &0.72 &3.6&0.13&0.68&1.9& $46.42$ & $2.0$ &1.1&$1.5$ \\
CO 10.999$-$0.902 & 3.4 & 0.26 &3.9&0.27&1.4 &0.79&$46.03$ &$1.6$ & 0.56&$2.0$ \\
CO 11.105$-$0.983 & 3.4 &0.29&1.8&0.067& 0.35 &0.19& $45.53$ & $1.6$  &0.18 & $0.26$ \\
CO 11.113$-$0.894 & 3.4 &0.73 &1.8&0.13&0.68 &0.47&$45.82$ &$4.0$ &0.14 & $2.3$ \\
CO 11.509$-$0.027 & 3.8&0.39 &1.8&0.076& 0.40 &0.25&$45.59$& $2.1$& 0.15 & $0.45$ \\
CO 11.771$-$0.075& 5.7&0.88 &2.2&1.7& 8.9 &0.80& $46.78$ & $5.7$ & 0.87&$3.9$ \\
CO 11.808+0.833 &3.0&0.50&3.6&0.40& 2.1 &1.3& $46.31$ & $2.7$ &0.63 & $4.9$ \\
CO 11.919$-$0.603 & 3.9&0.67 &3.6&0.25& 1.3 &1.7& $46.70$ & $1.8$ &2.3&$29$ \\
CO 11.949$-$0.039 & 4.4&0.99 &2.9&0.63& 3.3 &1.7& $46.92$& $3.3$ &2.1 &$23$ \\
CO 12.414+0.496 & 2.0 &0.38&2.2&0.36& 1.9 &0.37&$46.44$ & $1.7$ & 1.3 & $12$ \\
CO 12.433+0.505 &2.4&0.69 &1.8&0.50& 2.6 &0.45& $46.40$ &$3.8$ &0.55 & $9.7$ \\
CO 12.796$-$0.204 &3.8&1.3 &3.6&1.9&10 &3.4& $47.59$& $3.5$ & 9.2& $1.3\!\times\!10^{2}$ \\
CO 12.859$-$0.203 &5.0&1.6 &2.9&0.80& 4.2&2.7& $47.02$ & $5.4$ & 1.6 & $2.2\!\times\!10^{2}$ \\ 
CO 12.907$-$0.260  & 3.8 & 1.0&4.7&1.6& 8.5 &4.4& $47.75$ & $2.1$ & 22 & $27$\\
CO 12.927$-$0.204& 3.5 &0.58&2.2&0.27& 1.4 &0.56& $46.31$ & $2.6$ & 0.65 & $23$ \\
CO 13.367$-$0.378 &2.3 &0.36 &2.5&0.23&1.2&0.45& $46.35$ &$1.4$& 1.3 &$12$ \\
CO 13.461$+$0.241 &1.9 &0.64 &2.2&0.11& 0.58&0.62&$45.92$ &$2.9$ & 0.24 & $0.85$ \\
CO 13.968$-$0.388 & 2.0&1.1 &3.6&0.46& 2.4 &2.9&$46.97$ &$3.0$ & 2.6 & $59$ \\
CO 14.331$-$0.646 & 2.3 &0.32&6.5&0.78&4.1&2.7&$46.50$ & $2.0$ & 1.3&$9.1$\\
CO 15.946$-$0.386 &3.7&0.55&2.5&0.38&2.0 &0.69& $46.57$ &$2.2$ & 1.4 & $39$ \\
CO 16.134$-$0.553 &3.7&1.3 &9.0&13& 66 &21&$49.20$ & $1.2$& $9.5\!\times\!10^{2}$ &$<5.7$\tablenotemark{a} \\
CO 16.932+0.975& 1.9&0.60 &2.9&0.29& 1.5 &1.0& $46.58$ & $2.0$ &1.6&$2.6$ \\
CO 17.033+0.302& 2.1&0.47 &2.2&0.19& 0.98 &0.45& $46.15$ &$2.1$ & 0.56 & $2.2$\\
CO 17.625$+$0.165& 1.9&0.37 &2.2&0.23& 1.2 &0.36&$46.24$ & $1.6$ & 0.90 &$0.69$ \\
CO 18.298$-$0.398 & 2.7&1.2&5.4&1.0&5.5 &7.0&$46.90$& $5.3$ &1.2 &$12$ \\
CO 18.652$-$0.062& 3.6&1.0 &2.2&0.80& 4.2 &0.97&$46.78$& $4.5$ &1.1 &$11$ \\
CO 18.903$-$0.676& 4.7&1.1 &3.6&0.59&3.1 &2.9& $47.08$ & $3.0$ & 3.3 &$0.37$ \\
CO 22.838$-$0.476 & 5.0&3.1 &7.2&2.3& 12 &32& $48.27$ & $4.2$ &$37$& $12$ \\
CO 23.342$-$0.429  & 4.2&2.9&3.6&1.9& 10&7.5& $47.59$ &$7.9$ &4.1& $3.5$\\
CO 23.382$-$0.111& 6.0&0.83&2.5&0.15& 0.79 &1.0&$46.17$ &$3.3$ & 0.37 &$4.3$\\
CO 23.438$-$0.181 & 6.4&0.63 &5.4&0.57& 3.0 &3.7&$47.42$ & $1.1$ &20 &$25$ \\
CO 23.522$-$0.035 & 5.2&0.93 &4.3&1.7&8.7&3.4& $46.92$ &$5.1$ &1.4 &$5.0$ \\
CO 23.706+0.168 & 7.5&2.8&2.2&1.9& 10 &2.7& $47.16$ &$12$&1.0 &$76$\\
CO 25.642$-$0.350 &5.6&1.3 &3.2&0.29&1.5 &2.7& $46.66$ & $4.0$ & 0.95 &$0.53$ \\
CO 25.731$-$0.199 & 5.5 &1.5&3.2&1.0& 5.5&3.1& $47.22$ & $4.6$ & 3.0 &$16$ \\
CO 25.834$-$0.233 & 5.5&1.8&3.6&0.59& 3.1 &4.7& $47.43$ &$4.9$& 2.0&$0.76$ \\
CO 26.561$-$0.729 & 3.9&1.2 &10.0&2.1& 11 &24&$47.86$ & $2.5$ &24& $1.7\!\times\!10^{2}$\\
CO 26.646$-$0.825 &4.0&0.80&2.9&0.11& 0.56 &1.3& $46.15$ & $2.7$ & 0.43 &$3.4$ \\
CO 26.658+0.605 &5.2&0.96 &2.5&0.38&2.0&1.2& $46.57$ & $3.8$ & 0.82 & $0.50$\\
CO 30.208$-$0.095 & 7.0&2.0 &2.2&3.2&17 &1.9& $47.39$ & $8.9$ &2.3 & $1.8$ \\
CO 33.909+0.090& 6.0&1.7 &6.1&4.8& 25 &13& $47.63$ & $6.9$ & 5.2 & $59$ \\
CO 35.018+0.010 &4.7&1.1&3.2&0.65& 3.4 &2.3&$47.02$ & $3.4$ & 2.5&$1.1$ \\
CO 35.291+0.017 &6.4&0.73 &3.6&0.80& 4.2 &1.9&$47.21$ & $2.0$ &6.8& $0.88$\\
CO 36.386+0.786 & 4.4&0.98&3.2&0.27&1.4 &2.0&$46.63$ & $3.0$&1.2& $1.5$\\
CO 40.274$-$0.230 &4.9&1.1 &12.0&1.3&7.0&32&$47.88$ & $1.8$ & 35& $16$ \\
CO 43.159$-$0.029 & 0.61&0.26&4.3&0.041& 0.22 &0.97&$46.08$ & $0.57$ &3.3 & $1.6$\\
CO 43.169+0.010 & 0.73&0.31 &3.6&0.14& 0.72 &0.81& $46.45$&$0.85$ &2.7 &$2.3$ \\
CO 45.469+0.040 & 4.0&0.34 &7.9&1.1& 5.8 &4.2& $47.22$ & $1.1$ &13 & $36$ \\
CO 47.394+0.096 &2.8&0.41&2.5&0.048& 0.25 &0.51& $45.67$ & $1.6$ &0.24 & $3.0$ \\
CO 201.339+0.288 & 3.1&0.68&4.3&0.15& 0.81&2.5&$45.89$& $3.7$ &0.18&$1.7$ \\
CO 212.064$-$0.739 & 6.9 &1.1&3.6&1.7&9.2 &2.9&$47.04$ &$5.4$ &1.7&$22$ \\
CO 220.454$-$0.613 & 2.0&0.70 &4.3&0.19&1.0 &2.7&$46.16$ &$3.1$ &0.39 & $0.24$\\
\enddata
\tablenotetext{a}{Since no infrared counterpart is associated with CO 16.134--0.553, the 3$\sigma$ upper limit is presented here.}
\label{table2}
\end{deluxetable*}

\clearpage
\noindent
A ($l, b, V$) three-dimensional ${\rm^{12}CO}$ {\it J}=1--0 data FITS cube used in this study covers the longitudes ($10^{\circ}\!\!\! \leq l\! \leq$$50^{\circ}$, $198^{\circ}\! \!\leq \hspace{-0.9mm} l \hspace{-0.9 mm}\leq \!\!236^{\circ}$) with a latitudinal coverage of $|b|\!\leq \!1^{\circ}$ (Figure\ \ref{fig1}).\ The velocity coverage ranges from $\textit V_{\rm{LSR}}=-100\ \rm{to}+200\  \rm{km\ s^{-1}}$.\ The data were smoothed to a 1.3 km s$^{-1}$ velocity resolution, and resampled onto a $8\dotarc5\!\times8\dotarc5\!\times\!0.65\ \rm{km\ s^{-1}}$ grid for the first quadrant survey and a $1\dotarc5\!\times\hspace{-0.4 mm}1\dotarc5\!\times\!0.65\ \rm{km\ s^{-1}}$ grid for the third quadrant survey.\ The average rms noise levels in $T_{\rm MB}$ were $\sim\!\!1.47 \rm\ K$ for ${\rm^{12}CO}$ and $\sim\!\!0.69\rm\ K$ for ${\rm^{13}CO}$ in the first quadrant ($10^{\circ} \!\!\leq l \leq\!\!50^{\circ}$) survey, and $\sim\!\!1.10\! \rm\ K$ for ${\rm^{12}CO}$ and  $\sim\!0.56 \rm\ K$ for ${\rm^{13}CO}$ in the third quadrant ($198^{\circ} \!\leq l \leq \!236^{\circ}$) survey.
\subsection{Identification Scheme}
We identified broad-velocity-width molecular features (hereafter BVFs) with compact appearances by eye using the FITS viewing software ``takefits".\ The ``by eye'' identification often shows better performances than any automated schemes especially for faint features in crowded areas.\ In order to ensure the objectivity, we employed a systematic scheme for BVF identification (Figure\ \ref{fig2}).\ The identification scheme consists of a triple loop procedure.\ The outermost loop is the ``FITS file open" loop, where each step of this loop processes one FITS file.\ There are 89 FITS files in the FUGIN data archive, each of which covers 2$^{\circ}$ width in longitude.\ The inner double loop procedure searches for BVFs while scanning the Galactic longitudes and latitudes.\ These scans proceed from the minimum to maximum in both coordinates.

We set the BVF searching parameters for the FUGIN ${\rm CO}$ {\it J}=1--0 data as follows:\ maximum spatial size of 10\  $\rm{pc}$, minimum velocity width of 5  km s$^{-1}$ in full-width at zero-intensity (FWZI), and minimum main beam temperature 1\ K ($\sim \!1\sigma$).\ We also define the FWZI size parameter of each BVF by $\Delta S= D\rm{tan}(\sqrt{\Delta{\textit{l}\,}\Delta{\textit{b}}}$), where $\Delta l$ ($\Delta b$) is the FWZI extent of the BVF in $l$ ($b$) and $D$ is the distance to the parent cloud.\ Here we calculated the kinematic distance using the rotation curve described in \citet{Sofue2}.\ The Galactic constants are taken to be $R_{0}= 8.0\ \rm{kpc}$ and $V_{0}= 200\ \rm{km\ s^{-1}}$ in this paper (e.g., \citealt{Sofue2}, \citealt{Sofue}).\ For simplicity, the near-distance was employed in this study.\ Employing the far-distance gives an upward revision to $\Delta S$ by up to a factor of 10.\ The maximum size was set to be consistent with the definition of an HVCC \citep{Oka99}.\ The employed minimum velocity width of 5 km s$^{-1}$ is factor of 2$\sim$3 broader than those of `normal’ molecular clouds that have the same sizes as the BVFs in the Galactic disk. This criterion is roughly consistent with that for HVCCs in the CMZ, where normal molecular clouds have velocity widths of $\sim$20 km s$^{-1}$.
\begin{figure}[tbh]
\vspace{6 mm}
%\hspace{-3 cm}
\includegraphics[width=80 mm]{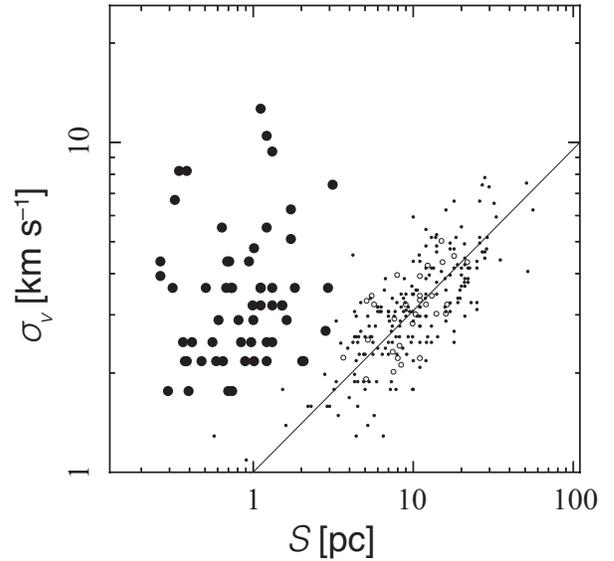}
%\centering
\caption{Plots of the velocity dispersion vs.\ size parameter.\ BVFs are shown with filled circles, while the Galactic disk clouds (\citealt{1987ApJ...319...730S}) are shown with dots.\ The open circles denote the cloud used to calculate $\Delta{S}/S$ and $\Delta{V}/\sigma_{V}$ ratios (see Appendix B).\ The solid straight line is $\sigma_{\textsl{v}}=\textit{S}^{\,0.5}\ \rm{km\ s^{-1}}$.}
\label{fig3}
\end{figure}

\begin{figure}[tbh]
\vspace{2 mm}
%\hspace{-3 cm}
\includegraphics[width=80 mm]{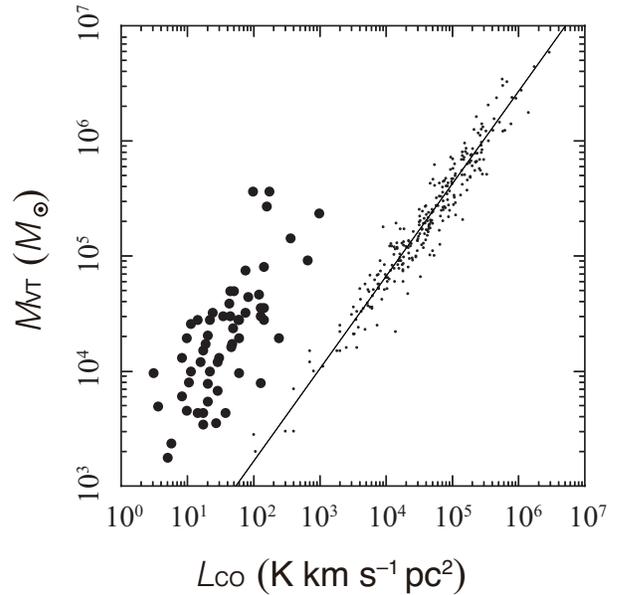}
%\centering
\caption{Plots of the virial theorem mass vs.\ CO luminosity.\ The filled circles and dots are the same as in Figure \!3.\ The solid straight line is $M_{\rm{VT}}=39(L_{\rm{CO}})^{0.81} M_\odot$.}
\label{fig4}
\end{figure}
\clearpage
\begin{figure}[tbh]
\vspace{6 mm}
\includegraphics[width=58 mm]{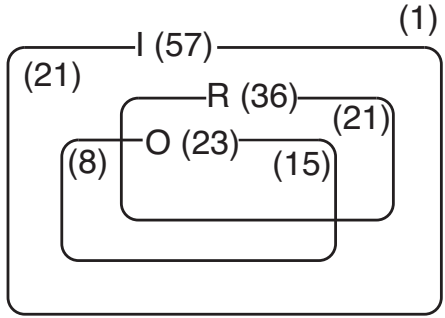}
\centering
\caption{Venn diagram showing the set relationship between the BVF categories:\,I (with infrared counterparts), R (with 10 GHz counterparts), O (molecular outflows).\ The number in the parentheses denotes that of the BVFs that belong to each set.}
\label{fig5}
\end{figure}

\section{Results} \label{sec:style}
\subsection{Identified BVFs}
By using the identification scheme described in the previous section, we successfully identified 58 BVFs in the FUGIN ${\rm^{12}CO}$ {\it J}=1--0 data.\ Note that some previously known broad CO line features, such as the W44 “Bullet” (\citealt{Sa13}) or the shocked clumps in SNRs (e.g.,\citealt{Re1999}; \citealt{Kil16}), have not been identified in the current survey.\ This is due to the fact that ${\rm^{12}CO}$ {\it J}=1--0 emission from these features is weaker than the rms noise of the FUGIN data ($\sim\!\!1.47 \rm\ K$; §2.1).\ A catalog of 58 BVFs is presented in Table \ref{table1} , and their images are shown in Appendix A. \ Each BVF in the catalog is listed with the following set of parameters: the $(l, b)$ location of the emission peak, velocity of the ``parent" molecular cloud, sizes along the $l$ and $b$ axes ($\Delta l$ and $\Delta b$), velocity width ($\Delta V$), peak temperature, and $^{13}$CO/$^{12}$CO intensity ratio.\ The sizes and velocity width are defined by FWZI.\

The $\it{l}$--$\it{b}$ distribution of velocity-integrated ${\rm^{12} CO}$ {\it J}=1--0 emission and the longitudinal distributions of BVFs in the first and third quadrant surveys are also shown in Figure \ref{fig1}.\ BVFs are widely distributed in the first quadrant, in which these are concentrated at longitudes from $l\!\!=10^{\circ}\,\rm{to}\,14^{\circ}$.\ In contrast to the first quadrant, where 55 BVFs were identified, only three BVFs were identified in the third quadrant.\ For the quantitative comparison with the Galactic disk molecular clouds, we calculated the size parameter ($S$) and velocity dispersion ($\sigma_{\textsl{v}}$) of each BVF, according to the method described in the Appendix B.\ The velocity dispersion ($\sigma_{\textsl{v}}$) of each BVF is plotted against its size ($S$) in Figure \ref{fig3}, with the same plot for the molecular clouds in the Galactic disk \citep{1987ApJ...319...730S}.\ The identified BVFs have sizes of 0.26 to 3.1 pc (average=0.93 pc) and velocity widths of 5--33 $\rm{km\ s^{-1}}$ (average=11 $\rm{km\ s^{-1}}$).\ Their small sizes and broad-velocity widths are natural consequences from the identification criterion of BVFs.\ The BVFs in the \textit{S}-$\sigma_{\textsl{v}}$ plot lie far above the fitted line for the Galactic disk clouds.\ One should note that employing the far-distances gives a rightward revision to the plot (see §2.2) bringing 10 BVFs on the disk cloud trend.\ The discrepancy in \textit{S}-$\sigma_{\textsl{v}}$ plots between BVFs and disk clouds indicates that they have different properties.\ 
\subsection{Physical Parameters}
Table \ref{table2} summarizes the physical parameters of identified BVFs.\ The cloud mass $M$ was derived from the CO luminosity ($L_{\rm{CO}}$) by,
\begin{equation}
M=\mu m_{\rm{H_{2}}}X_{\rm{CO}}L_{\rm{CO}}
\end{equation}
where $\mu$ is the mean molecular weight (1.38), $m_{\rm{H_{2}}}$ is the mass of molecular hydrogen ($3.34\!\times\!10^{-24}$ g) and $X_{\rm{CO}}$ is the CO-to-$\rm{H}_{2}$ conversion factor ($2.0\!\times\!10^{20}\ \rm{cm^{-2}(K\ km\ s^{-1})^{-1}}$;\,\citealt{Bolatto57}).\ The total CO luminosity,
\begin{equation}
L_{\rm{CO}}=D^{2}\int\int\int T_{R}^{*}(\rm{CO})\ d\!\textit{V}d\textit{l}d\textit{b}
\end{equation}
was calculated by integrating $T_{R}^{*}$ over all pixels of each BVF.\ Where $D$ is the distance to the cloud.\

The kinetic energy ($E_{\rm{kin}}$) and dynamical timescale ($t_{\rm{dyn}}$) were derived with the formulae:
\begin{equation}
E_{\rm{kin}}= \frac{3}{2}M\sigma_{V}^{2}\\
\end{equation}\
\begin{equation}
t_{\rm{dyn}}= \frac{S}{\sigma_{V}}.
\end{equation}

The virial theorem mass ($M_{\rm{VT}}$) was calculated with
\begin{equation}
M_{\rm{VT}}=3f_{p}\frac{S\sigma_{V}^{2}}{G}
\end{equation}
where $f_{p}$ is a projection factor and $G$ is the Gravitational constant.\ To keep the consistency with the previous studies, we employed $f_{p}$= \!2.9 \citep{1987ApJ...319...730S}.\ Figure\ \ref{fig4} shows the virial theorem mass--CO luminosity relations for BVFs and disk clouds.\ This clearly shows that BVFs belong to the populations different from normal disk clouds.\ Note that employing the far-distances gives an upper-rightward revision to the plot bringing seven BVFs on the disk cloud trend.

\subsection{Counterparts in other Wavelengths}
In order to search for the counterparts of BVFs in the other wavelengths, we examined the existing infrared and radio continuum data sets.\ We used the data from the Two Micron All Sky Survey (2MASS) at \textit{J} (1.25 \micron), \textit{H} (1.65 \micron), and \textit{K} (2.17 \micron) bands, Wide-field Infrared Survey Explorer (WISE) at 3.4 and 4.6 \micron\ bands, and AKARI at 60--140\ \micron.\ All the infrared data were obtained from the \textsl{skyview} website (\citealt{McGlynn98}; \citealt{McLean98}).\ We referred to the 10 GHz continuum data obtained with the NRO 45 m telescope \citep{1987PASJ...39..709H} and VLA 1.4 GHz data obtained from the \textsl{skyview} website.\ These comparisons may be useful to check the relation of BVFs to star forming regions and supernova remnants.\ 

We also classified the identified BVFs into three categories.\ The first category, named ``I", indicates BVFs with counterparts in infrared maps.\ The second category,``R", indicates BVFs with counterparts in radio continuum maps.\ The third category,\ ``O", indicates BVFs that have already been identified as molecular outflows in the previous literature.\ Each BVF type in Table \ref{table1} is denoted by a combination of these classifications.\ As a result, we found 36 BVFs with counterparts in both of the infrared and radio continuum maps, and 15 of them are considered as molecular outflows in the literature \citep{Cyagano2008, 2017ApJ...729..124C, 2011MNRAS.415L..49C, 2012AZh...56..731C, Maud15, Li2018, Yang2018, 2019ApJ...886L...4Z}.\ Twenty-one BVFs have counterparts only in infrared maps, and eight of them are described as molecular outflows in the literature (\citealt{Lee13}; \citealt{Fro15}; \citealt{Maud15,Maud18}; \citealt{Chen16}; \citealt{2019MNRAS.485.1775I}).\ Figure \ref{fig5} shows the set relationship between the BVF categories as a Venn diagram.\ Some BVFs are found at positions near well-known massive star forming regions, namely, W33 and W31.\ On the other hand, no BVF was found to be associated with cataloged supernova remnants \citep{2019AAP...40...36G}.\ We also examined the data from the HI, OH, recombination line survey of the Milky way with the VLA\ \citep{Be16, Wang2020}.\ We found that six BVFs, CO 23.382--0.111, CO 25.834--0.233, CO 26.561--0.729, CO 26.658+0.605, CO 30.208--0.095, and CO 45.469+0.040 are possibly associated with HI counterparts.\ These HI counterparts are spatially more extended than BVFs by a factor of 1$\sim$2, and their velocity extents are less definite because of widespread HI emission from unrelated gas.\ It was suggested in previous studies that high-velocity HI gas features can be produced by stellar feedback, such as outflows from young stellar objects (YSOs), stellar winds from massive stars, and supernova explosions\ \citep{2007ApJS.173.85K}.\ Thus, it is likely that these six BVFs have also been formed by protostellar outflows, although the other feedback processes can not be ruled out.

%\clearpage
\section{Discussion}
\subsection{BVFs with Infrared Counterparts}
\begin{figure*}[tbh]
\vspace{2 cm}
\includegraphics[width=180 mm]{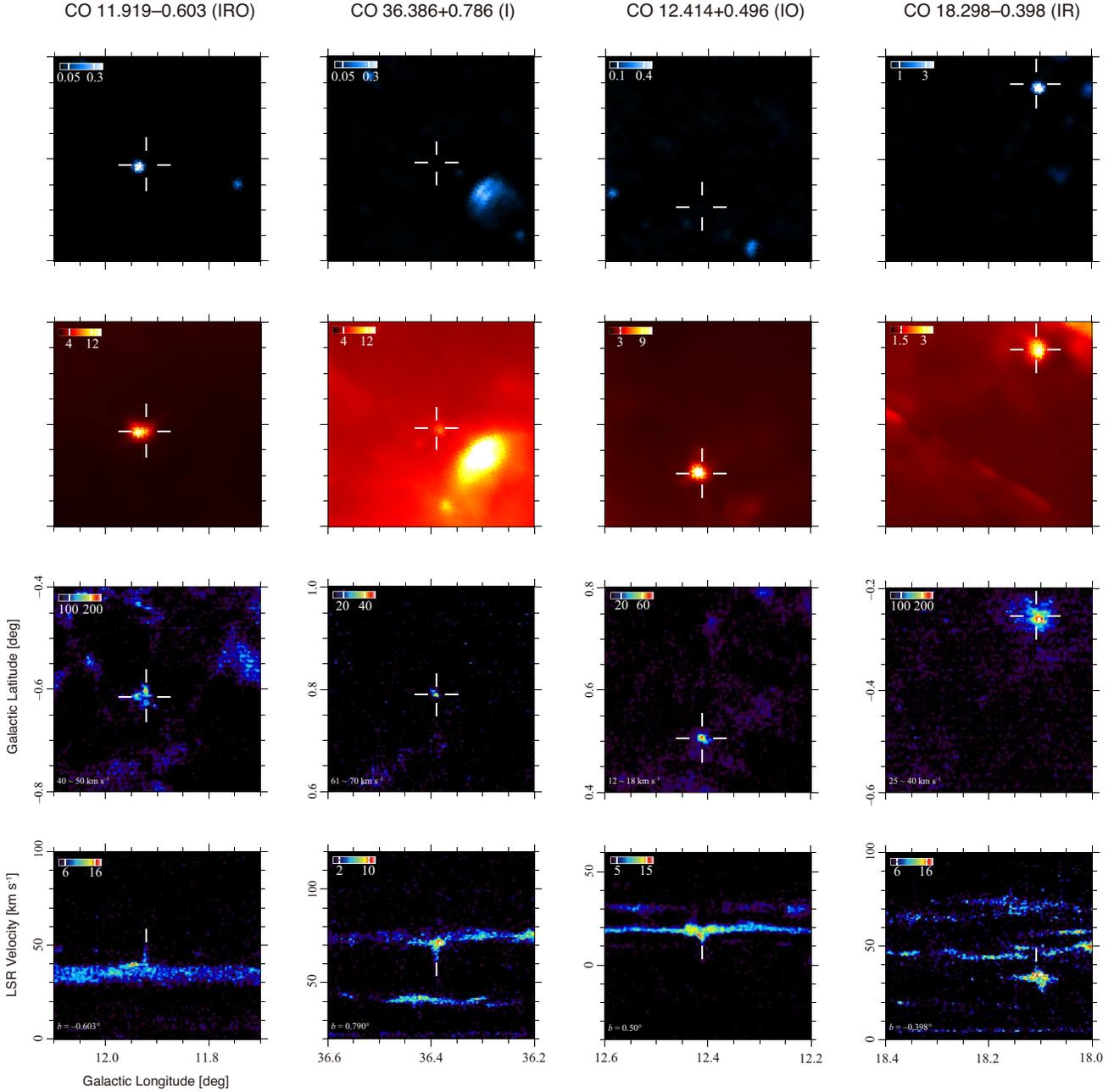}
\centering
\caption{Maps of radio continuum, far-infrared, and CO {\it J}=1--0 emission toward four BVFs representing each type.\ Panels in each column show: VLA 1.4 GHz image, AKARI WIDE-S (60--110\ \micron) image, map of velocity-integrated CO {\it J}=1--0 emission, and longitude--velocity map of CO {\it J}=1--0, from top to bottom.\ The intensity unites are, Jy/beam, Jy/str, K km s$^{-1}$, and K, respectively.\ White markers show the position of BVFs.}
\label{fig6}
\end{figure*}

Figure \ref{fig6} shows the CO, infrared, and radio continuum maps of typical BVFs in each type.\ As shown in the previous section, most of the identified BVFs have their infrared counterparts (categorized into I-type).\ Twenty-three of them have been identified as molecular outflows.\ Fifteen of the I-type BVFs show double-sided high-velocity wings (seventh column in Table \ref{table1}).\ In addition, some BVFs coincide with the position of young stellar YSOs (e.g.,\  \citealt{Felli02}; \citealt{Robi08}; \citealt{Urq09, Urq15}; \citealt{Saral15, Saral17}).\ From these properties, together with the inclusion relationship between the BVF categories (Figure \ref{fig5}), it is most likely that all of the I-type BVFs may be bipolar molecular outflows from YSOs.\ 

The parameter ranges of I-type BVFs are, $S$=0.26--3.1 pc, $\Delta{V}$=5--33 $\rm{km\ s^{-1}}$, $M$=22--4500 $M_\odot$, $t_{\rm{dyn}}$=$5.7\!\times10^{4}$--$1.2\times10^{6}$ yr, \!and $E_{\rm{kin}}$=$3.4\times10^{45}$--$1.8\times10^{48}$ erg.\ These are reasonably overlapped with those of protostellar outflows (e.g., \citealt{Beuther2002}; \citealt{Maud15}; ; \citealt{Li2018}; \citealt{Zhang2020}).\ The absence of less energetic BVFs may be a result of selection effect from different survey sensitivities.\ The non-I-type BVF, CO 16.134--0.553, which have a kinetic energy well greater than the range of protostellar outflows, will be discussed in §4.2.\ We also checked the color of the associated infrared sources (e.g., \citealt{Yama10, Yama18}).\ Most infrared sources associated with BVFs have more pronounced emission in the shorter wavelengths than in the longer wavelengths.\ This behavior is common for YSOs.

\begin{figure}[h]
\vspace{3 mm}
\includegraphics[width=80 mm]{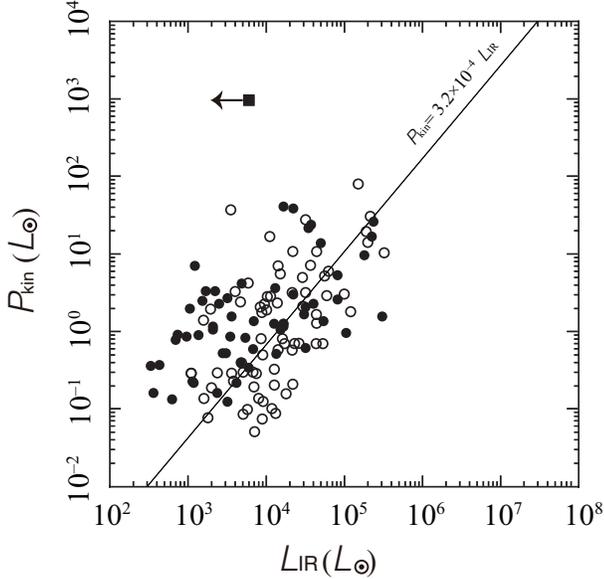}
%\centering
\caption{Plots of power ($P_{\rm kin}$) against the source infrared luminosity ($L_{\rm IR}$).\ The open circles are the outflow sources in \citet{Maud15}.\ The solid circles are BVFs.\ The filled square is CO 16.134--0.553.\ The solid straight line, which only uses data from \citet{Maud15} shows the best-fit line, $P_{\rm kin}=3.2\times10^{-4}L_{\rm IR}$.}
\label{fig7}
\end{figure}

\begin{figure*}[tbh]
\vspace{-0.2 cm}
\includegraphics[width=120 mm]{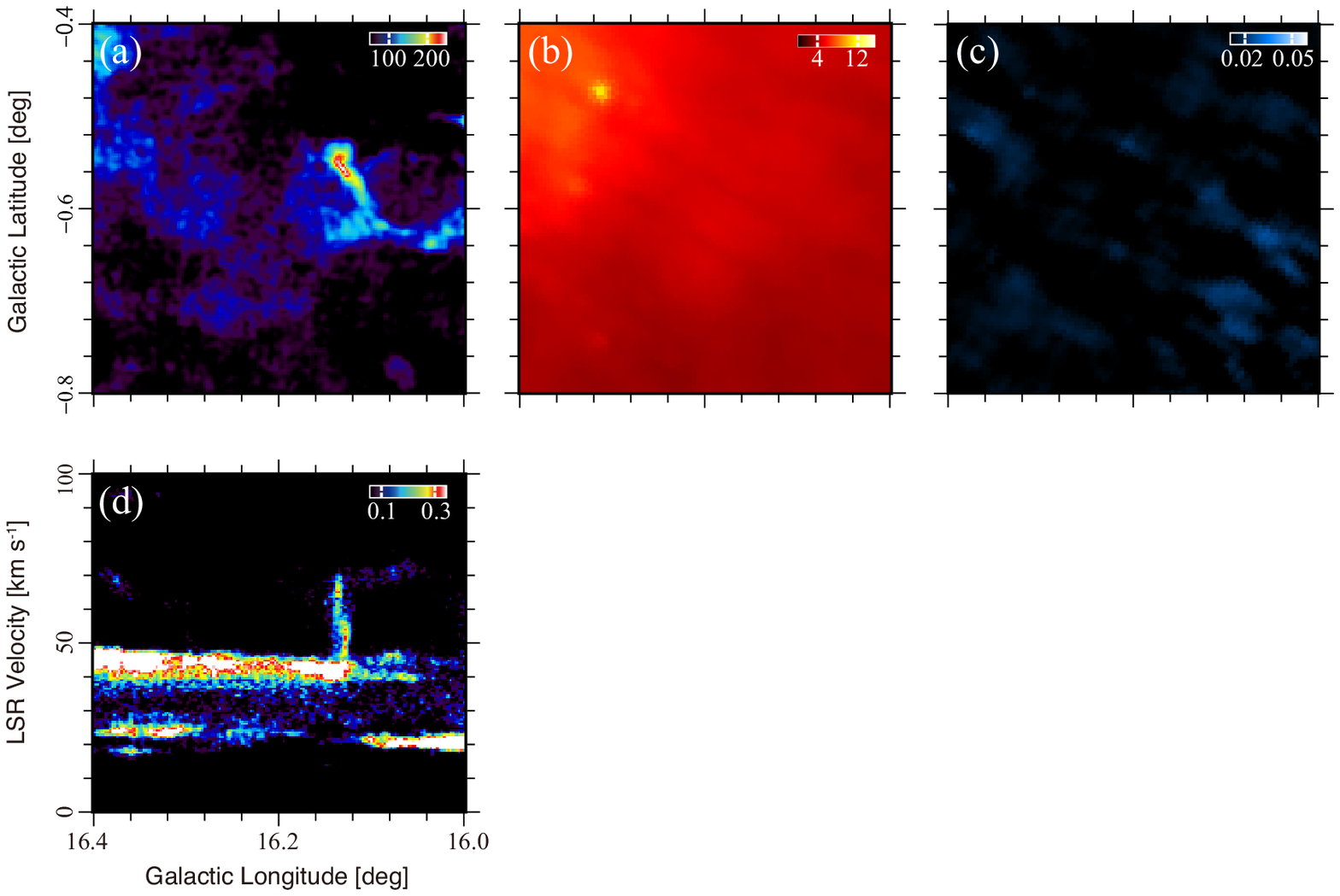}
\centering
\caption{CO, infrared, and radio continuum maps of CO 16.134--0.553 : (a) Map of the velocity-integrated CO {\it J}=1--0 line emission. The velocity range for the integration is from $\textit V_{\rm{LSR}} = 45\ \rm{to}\ 70\ \rm{km\ s^{-1}}$.\ The intensity unit is K $\rm{km\ s^{-1}}$.\ (b) AKARI WIDE-S (60--110\ \micron)\ image of the same field.\ The intensity unit is Jy/str.\ (c) VLA 1.4 GHz image of the same field.\ The intensity unit is Jy/beam.\ (d) Map of the latitude-integrated CO {\it J}=1--0 line emission. The latitude range for integration is from $b=-0\dotdeg52\ \rm{to} -0\dotdeg58$.\ The intensity unit is K.}
\label{fig8}
\end{figure*}
Then we calculated the kinetic power [$P_{\rm {kin}}=E_{\rm{kin}}/(S/\sigma_{\textsl{V}})$] of each BVF and infrared luminosity $L_{\rm IR}$ of its counterpart.\ The infrared luminosity was calculated from the fluxes in the AKARI/FIS all-sky survey point source catalog (\citealt{Yama10, Yama18}) by using the following formula presented in \citealt{2016A&A...595C...1S} 
:
%\vspace{-0.3 cm}
\begin{eqnarray}
L_{\rm{AKARI}}^{\rm{3bands}}= \Delta \nu(N60)L_{\nu}(65 \ \rm \mu m)+ \nonumber \\
\Delta \nu(\rm{WIDE\mathchar`-S})\textit{L}_{\nu}(90 \ \rm \mu m)+ \nonumber \\
\Delta \nu(\rm{WIDE\mathchar`-L})\textit{L}_{\nu}(140 \ \rm \mu m),
\end{eqnarray}\
where
\begin{align*}
\Delta \nu(N60) = 1.58\times10^{12}\,[\rm{Hz}] \\
\Delta \nu(\rm{WIDE\mathchar`-S}) = 1.47\times10^{12}\,[Hz] \\
\Delta \nu(\rm{WIDE\mathchar`-L}) = 0.831\times10^{12}\,[Hz].
\end{align*}

\noindent
The luminosity at each wavelength was calculated from the flux density by $L_{\nu}=4\pi D^{2} F_{\nu}$.\ The kinematic distance ($D$) was derived from the LSR velocity of the parent cloud.\ Note that $L_{\rm IR}$ is proportional to $D^{2}$ while $P_{\rm kin}$ depends linearly on $D$.\ Figure \ref{fig7} shows the $L_{\rm IR}$--$P_{\rm kin}$ plot of BVFs/IR counterparts with that of outflow sources (\citealt{Maud15}).\ The $L_{\rm IR}$--$P_{\rm kin}$ plot of I-type BVFs well overlaps with that of outflow sources.\ Employing the far-distances gives an upper-rightward revision to the BVF plot, which brings no significant change to their $L_{\rm IR}$--$P_{\rm kin}$ loci relative to the outflows.\ The slope of I-type BVFs is slightly flatter than that of the outflows, having higher $P_{\rm kin}$ at $L_{\rm IR}$ $<$ 10$^{4}$ $L_{\rm sun}$.\ This discrepancy may demonstrate a result of selection effects: difference between CO--selected and IR--selected outflows.\ Despite the slight discrepancy, it may be reasonable to say that the $L_{\rm IR}$-$P_{\rm kin}$ behavior of these two sets follows the same trend.\ The results clearly show that the I-type BVFs follow the same trend as outflow sources in the $L_{\rm IR}$--$P_{\rm kin}$ plane.\ These facts suggest that these two sets belong to the same population of celestial objects---protostellar outflows.\

The number and distribution of protostellar outflows revealed by the H$_{2}$ 2.122\ \micron\ line survey could be a challenge to our interpretation.\ There are about a hundred of H$_{2}$-detected outflows in the Serpent+Aquila region (\citealt{Froe2012}), while only 13 I-type BVFs were found in the same region.\ In addition, two BVFs (CO 26.561--0.729 and CO 26.646--0.825) do not have their H$_{2}$ counterparts.\ These discrepancies may be results of selection effects: (1) the near-infrared H$_{2}$ line survey is more sensitive to small, less energetic outflows, and (2) the millimeter-wave CO survey does not suffer from interstellar extinction.
\subsection{BVF without an Infrared Counterpart}
Only one BVF, CO 16.134--0.553, appeared to have no counterparts in the other wavelengths.\ In the CO {\it J}=1--0 maps, this BVF shows an elongated spatial structure with a spatial size of $3\times4$ pc$^2$, and it has a single-sided high-velocity wing with a particularly broad-velocity width ($\Delta V \!\simeq\! 30 $ km s$^{-1}$; Figure \ref{fig8}).\ The kinematic distance was estimated from the LSR velocity of the parent cloud, $V_{\rm LSR}\!\simeq\!45$ km s$^{-1}$, to be $D\!=\!3.7$ kpc, which corresponds to the line-of-sight location of the Norma arm \citep{2019PASJ...tmp...50T}.\ The kinetic power of CO 16.134--0.553 is as high as $\sim\!\!10^3 L_{\odot}$, while the infrared luminosity is less than $\sim\!\!10^4 L_{\odot}$.\ These values differ significantly from the $L_{\rm IR}$--$P_{\rm kin}$ trend of molecular outflows (Figure \ref{fig7}).\ The compact appearance and broad-velocity width of CO 16.134--0.553, as well as the absence of a luminous counterpart, remind us of a population of peculiar molecular clouds found in the Galactic center, HVCCs.\ We interpret that CO 16.134--0.553 may be an analog of an HVCC in the disk part of our Galaxy.\ Only one HVCC analog was detected in the $40^{\circ}\!\times2^{\circ}$ area, yielding a number density of 0.01 deg$^{-2}$.\ On the other hand, the number density of HVCC in the CMZ is 122 deg$^{-2}$.\ This sharp discrepancy may indicate the approximate uniqueness of HVCCs to the CMZ. Detailed discussions about the nature and origin of CO 16.134--0.553 based on the follow-up observations will be presented in a forthcoming paper (\citealt{Yoko}).

\section{Summary}
We have performed an unbiased search for compact broad-velocity-width molecular features (BVFs) in the Galactic plane by using the FUGIN ${\rm^{12}CO}$ {\it J}=1--0 survey data.\ The purpose of our survey was to examine the universality of HVCCs in the Milky Way.\ Our results are summarized as follows:\

1. By employing specific criteria, i.e., a spatial size smaller than 10 pc and a velocity width larger than 5  km s$^{-1}$, we identified 58 BVFs in total.\

2. Most of the BVFs were found in the first quadrant ($10^{\circ}\! \leq \!l\! \leq \!\!50^{\circ}$) of the Galactic plane, and only three BVFs were found in the third quadrant ($198^{\circ}\! \leq \!l\!\leq 236^{\circ}$).\

3. All but one of the identified BVFs have infrared counterparts, which show the same $L_{\rm IR}$--$P_{\rm kin}$ trend with protostellar outflows.\ It is most likely that they are molecular outflows driven by YSOs.\

4. One BVF without a luminous counterpart has a huge kinetic power ($P_{\rm kin}\sim\!10^3 L_{\odot}$) that significantly departs from the $L_{\rm IR}$--$P_{\rm kin}$ trend of protostellar outflows.\ This CO 16.134--0.553 may be an analog of an HVCC in the Galactic center.\

This study provides a list of protostellar outflows, many of which have been unrecognized previously.\ The list serves as an important compilation of essential targets for studies of YSOs.\ The detection of CO 16.134--0.553 will contribute to the understanding of the similar in the Galactic center molecular clouds with extremely broad-velocity widths, which might be related to the evolution of disk galaxies.

\acknowledgments
This study is based on observations at the Nobeyama Radio Observatory (NRO).\ The NRO is a branch of the National Astronomical Observatory of Japan, National Institutes of Natural Sciences.\ In addition, this research is based on observations with AKARI, a JAXA project with the participation of ESA.\ This paper makes use of the FUGIN data\footnote{\url{https://nro-fugin.github.io/release/}}.\ The data were retrieved from the JVO portal\footnote{\url{http://jvo.nao.ac.jp/portal/}} operated by ADC/NAOJ.\ We acknowledge the use of NASA's \textsl{skyview} facility\footnote{\url{http://skyview.gsfc.nasa.gov/}} located at the NASA Goddard Space Flight Center.

\appendix
\section{The CO images and $(p-v)$ diagrams of all 58 BVFs}
We present the CO images and longitude--velocity diagrams of all 58 BVFs in Figure \ref{fig9}.
\section{The spatial size and velocity dispersion of BVFs}
The spatial size and velocity width of BVFs in this study are defined in full-widths at zero-intensity (FWZI).\ For quantitative analyses, however, we need to have the spatial and velocity dispersions.\ To derive these dispersions, we examined the correspondence between those and FWZI sizes by using well-defined molecular clouds in the Galactic disk.\ We referred to \citet{1987ApJ...319...730S} and picked up 30 isolated clouds,\ \footnote{Cloud numbers in \citet{1987ApJ...319...730S} are as follows : 15, 16, 18, 20, 23, 24, 26, 29, 31, 32, 34, 45, 56, 62, 64, 78, 111, 113, 117, 118, 123, 131, 132, 135, 136, 137, 138, 139, 141, 142.} which also appear in the FUGIN survey coverage.\ The spatial size of clouds in \citet{1987ApJ...319...730S} was defined by the size parameter: 
\begin{equation}
S= D {\rm tan}(\sqrt{\sigma_l \sigma_b })
\end{equation}
where $D$ is the distance to the cloud and $\sigma_{\rm{x}}$ is the dispersion in the direction of ``x".\ Explicitly, $\sigma_{\rm{x}}=(\overline{\rm{x}^{2}}-\overline{\rm{x}}^{2})^{1/2}, \overline{\rm{x}}=\sum{T_{\rm{x}}}/\sum{T}, \rm{x}=\textit{l, b,V}.$\ We estimated the FWZI sizes of picked up clouds in terms of the Galactic longitude ($\Delta l$), latitude ($\Delta b$), and velocity ($\Delta V$) by using the FUGIN CO {\it J}=1--0 data.\ The FWZI size parameter was calculated by 
\begin{equation}
\Delta S= D\rm{tan}(\sqrt{\Delta{\textit{l}}\,\Delta{\textit{b}}})
\end{equation}\
Figure \ref{fig10} shows the plots of $S$ versus $\Delta S$ and $\sigma_{V}$ versus $\Delta V$ for the picked up 30 clouds.\ The linear least-squares fittings (without the 0-th order term) gave the relations
\begin{eqnarray}
%\begin{equation}
\Delta S  =  (2.13\pm 0.39) S\\
\Delta V  =  (2.79\pm 0.24) \sigma_{V}
\end{eqnarray}\
%\end{equation}\
By using these relations, we calculated $S$ and $\sigma_{V}$ of the identified BVFs, which were used to calculate their physical parameters such as the expansion time, mass, and kinetic energy.

\begin{figure}[tbh]
\includegraphics[width=145 mm]{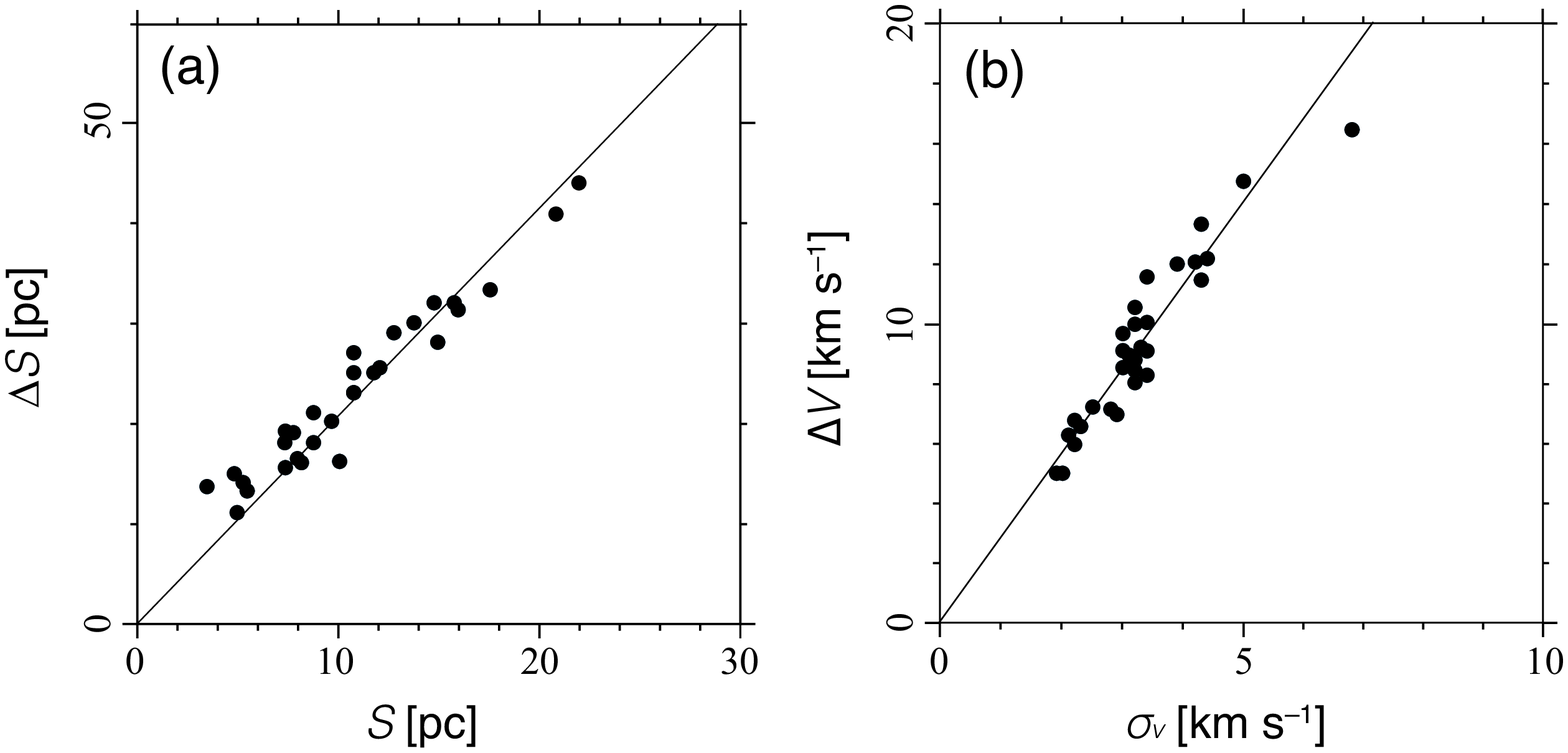}
\setcounter{figure}{9}
\centering
\caption{(a) Plot of the size parameter versus FWZI size for the 30 molecular clouds in \citet{1987ApJ...319...730S}.\ The solid straight line shows the best-fit line, $\Delta S = 2.13 S$.\ (b) Plot of the velocity dispersion versus FWZI velocity width for the 30 molecular clouds in \citet{1987ApJ...319...730S}.\ The solid straight line shows the best-fit line, $\Delta V = 2.79 \sigma_{V}$.}
\label{fig10}
\end{figure}
\clearpage

\begin{figure*}[tbh]
\setcounter{figure}{8}
\vspace{1.5 cm}
\includegraphics[width=130 mm]{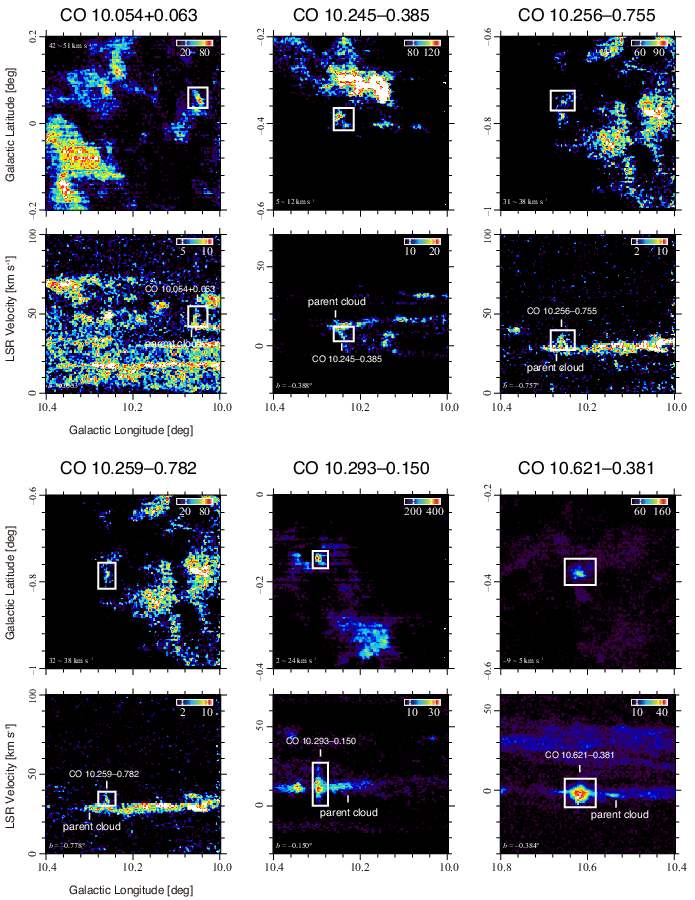}
\centering
\caption{The CO images and longitude--velocity diagrams of all 58 BVFs.\ The intensity unit of CO images is K $\rm{km\ s^{-1}}$.\ The intensity unit of longitude--velocity diagrams is K.\ The white rectangles show the $\Delta{l}\times\Delta{b}\times\Delta{V}$ extent of BVFs.}
\label{fig9}
\end{figure*}

\begin{figure*}[tbh]
\setcounter{figure}{8}
\vspace{-0.2 cm}
\includegraphics[width=130 mm]{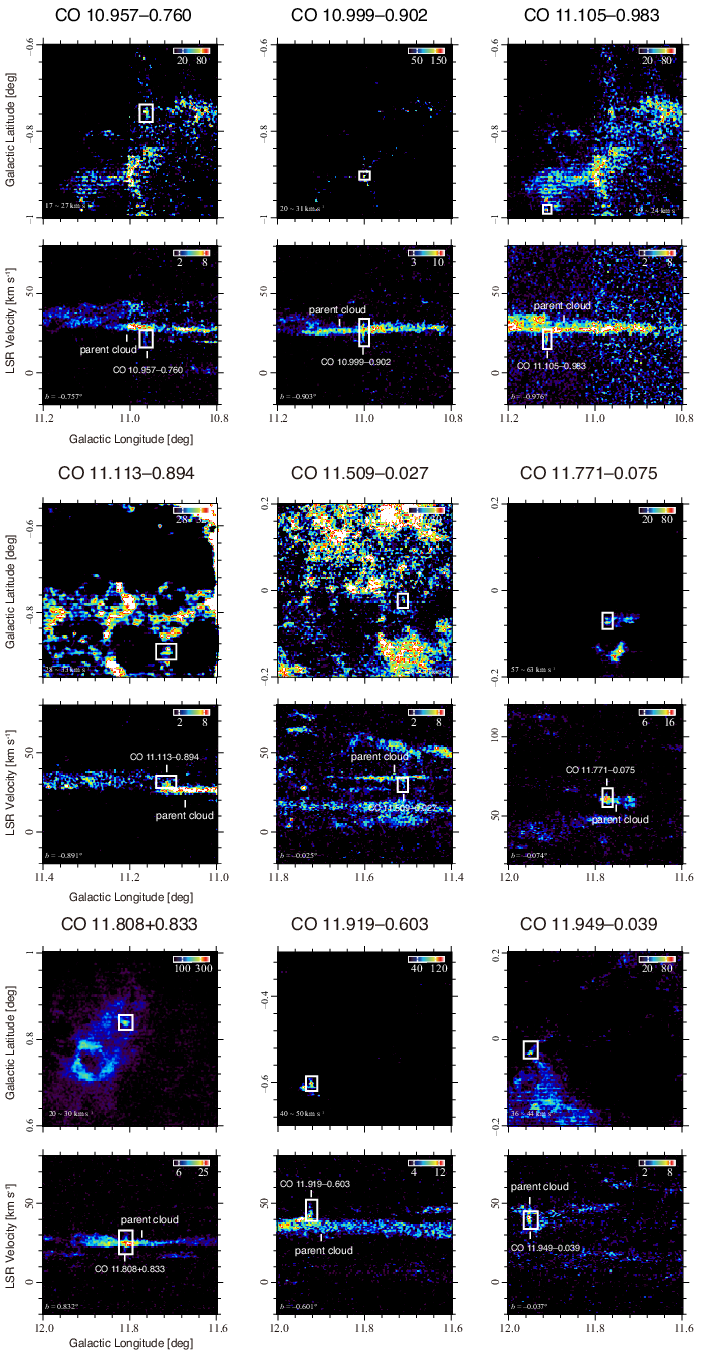}
\centering
\caption{Continued}
\label{fig9}
\end{figure*}

\begin{figure*}[tbh]
\setcounter{figure}{8}
\vspace{-0.1 cm}
\includegraphics[width=130 mm]{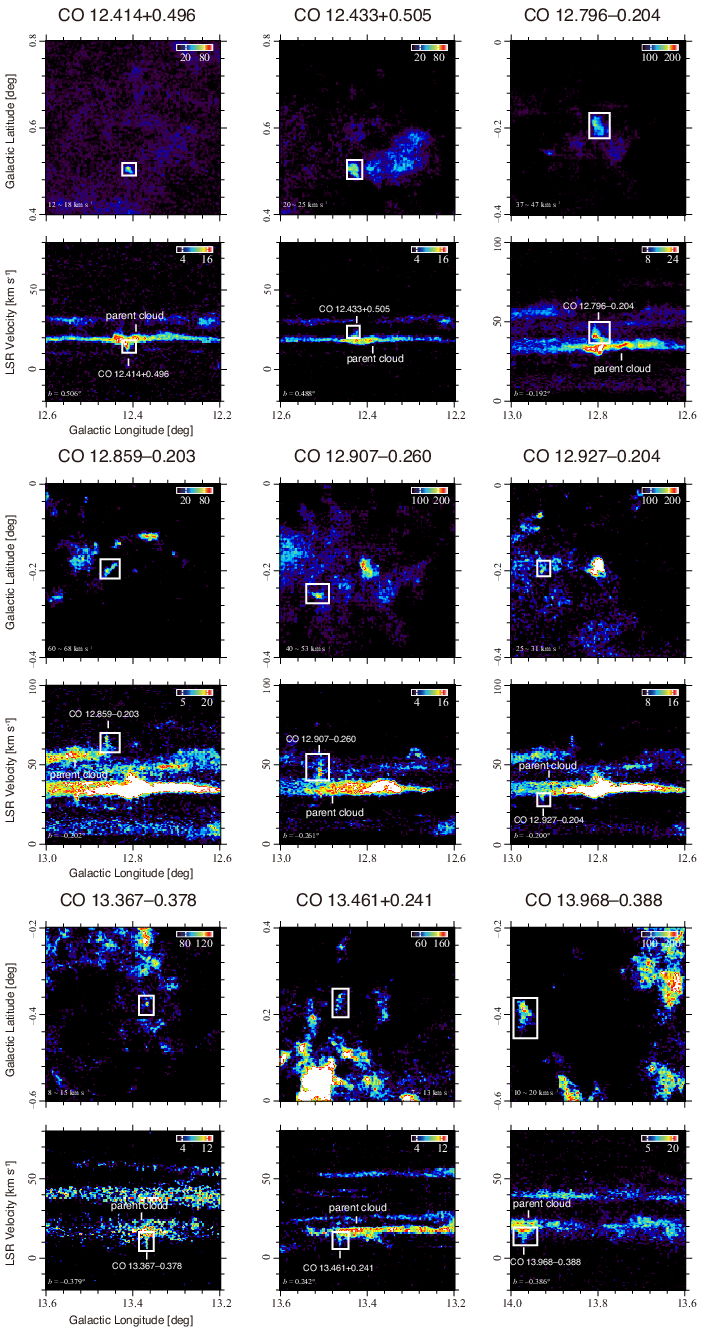}
\centering
\caption{Continued}
\label{fig9}
\end{figure*}

\begin{figure*}[tbh]
\setcounter{figure}{8}
\vspace{-0.1 cm}
\includegraphics[width=130 mm]{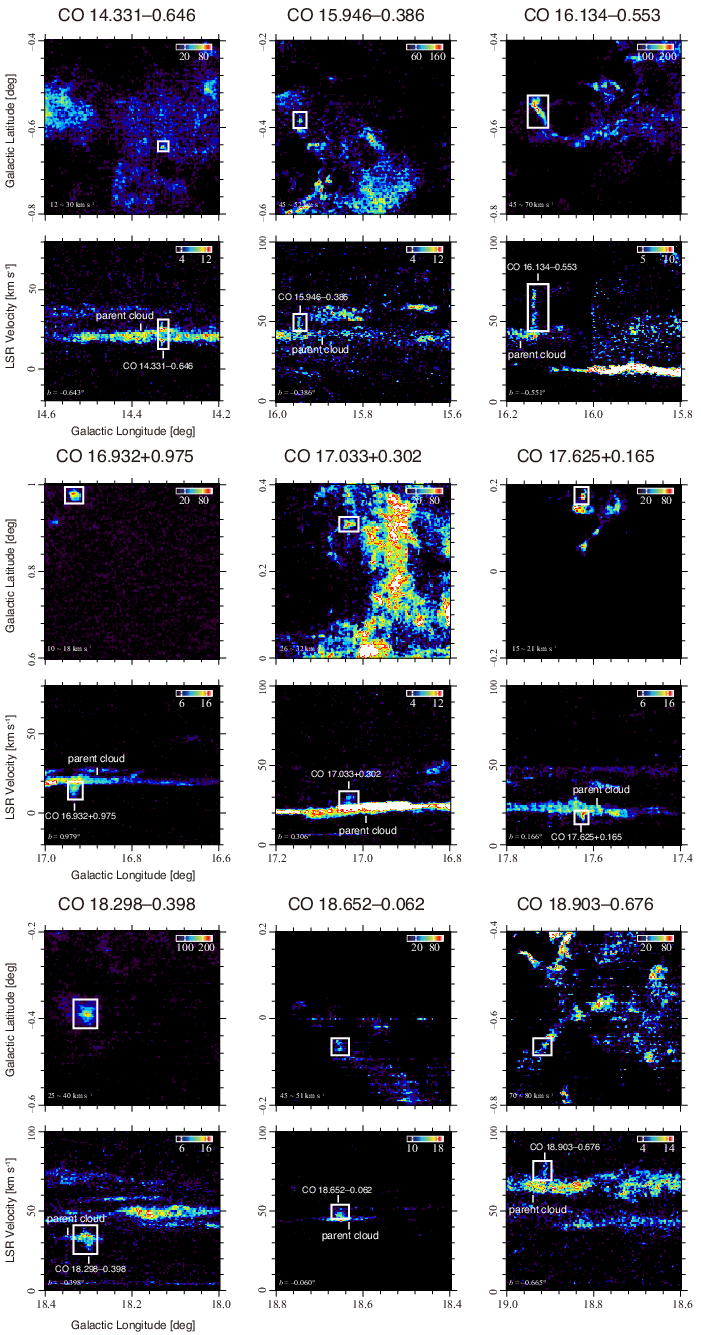}
\centering
\caption{Continued}
\label{fig9}
\end{figure*}

\begin{figure*}[tbh]
\setcounter{figure}{8}
\vspace{-0.1 cm}
\includegraphics[width=130 mm]{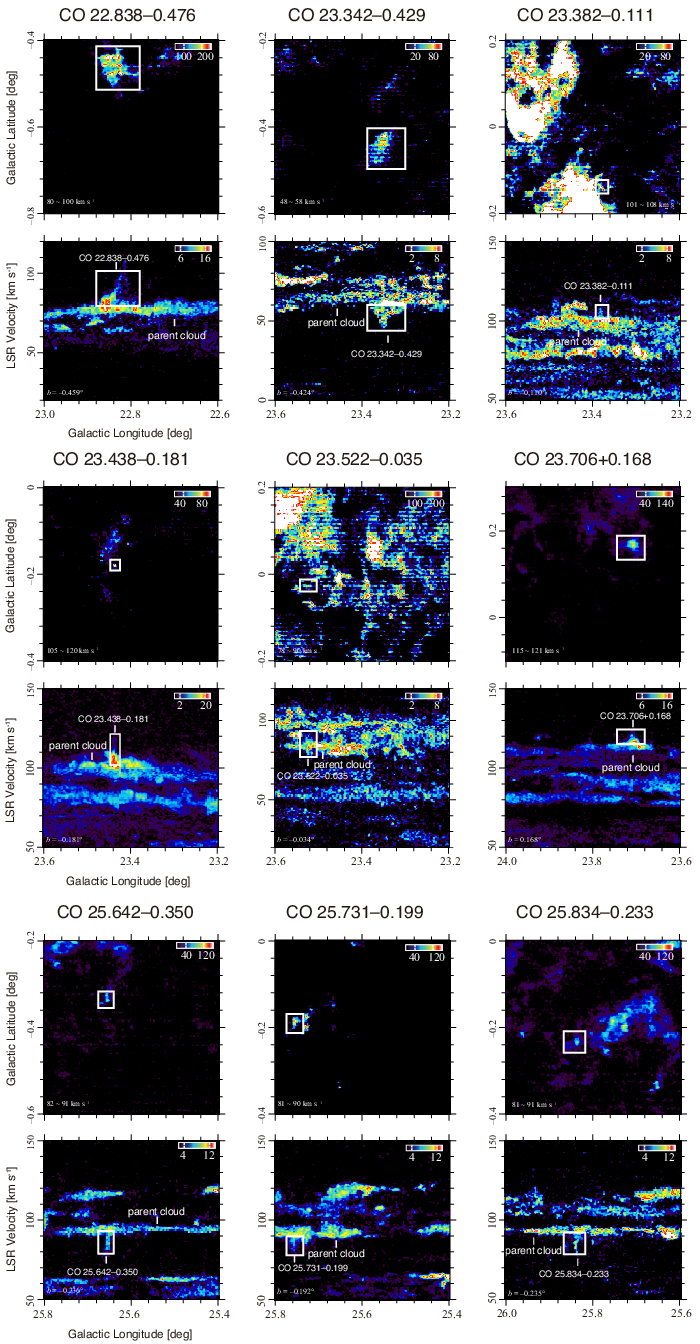}
\centering
\caption{Continued}
\label{fig9}
\end{figure*}

\begin{figure*}[tbh]
\setcounter{figure}{8}
\vspace{-0.15 cm}
\includegraphics[width=130 mm]{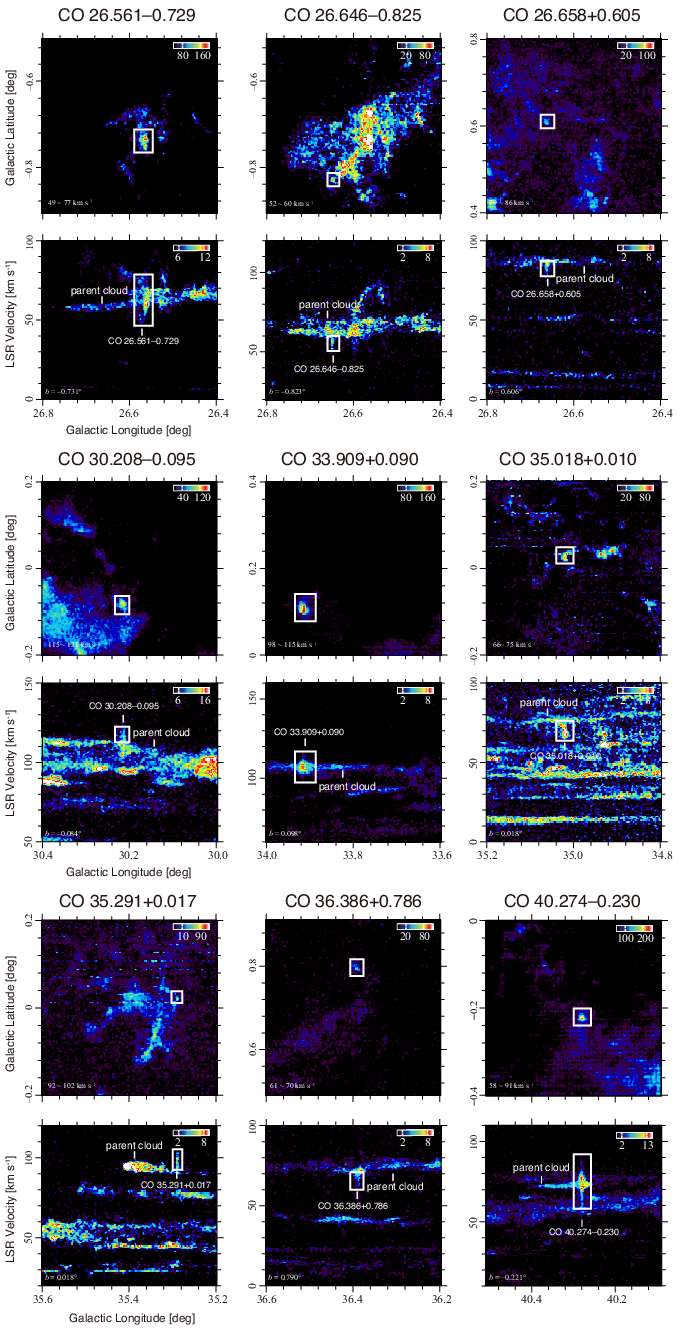}
\centering
\caption{Continued}
\label{fig9}
\end{figure*}

\begin{figure*}[tbh]
\setcounter{figure}{8}
\vspace{-0.15 cm}
\includegraphics[width=130 mm]{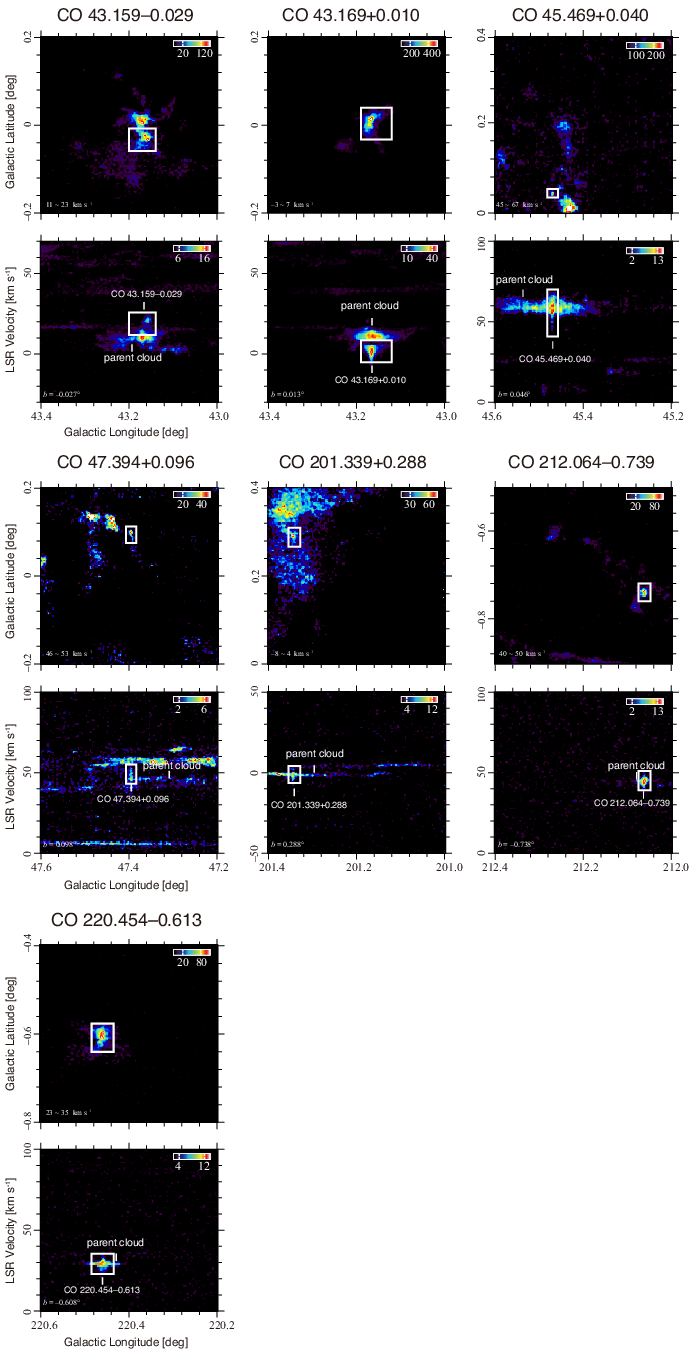}
\centering
\caption{Continued}
\label{fig9}
\end{figure*}
\end{document}